# Fast nuclear rotation and octupole deformation

W. Urban[1], J.C Bacelar[2] and J. Nyberg[3]

[1] Institute of Experimental Physics, Warsaw University, Warsaw, Poland,
[2] KVI, University of Groningen, Groningen, The Netherlands,
[3] NBI Riso, Roskilde, Denmark

The $^{150}$Sm nucleus has been studied to high spins in a measurement of $\gamma$ radiation following the $^{136}$Xe($^{18}$O,4n)$^{150}$Sm, compound-nucleus reaction at beam energy of 76 MeV. The measurement was performed at NBI Risø using the NORDBALL array. Alternating parity, s=+1 band in $^{150}$Sm has been observed up to spin I=22. This band is crossed by two aligned bands, corresponding to a reflection-symmetric shape. After the second crossing the s=+1 band ends abruptly, suggesting that the octupole shape vanishes in $^{150}$Sm above spin I=22, as predicted by calculations. Other explanations, assuming continuation of the s=+1 band past the two alignments are also discussed.

PACS numbers: PACS 23.20.Lv, 21.10.Re, 27.60.+j

## 1. Introduction

Enhanced octupole interactions, which may cause octupole deformation, are expected in nuclei where opposite-parity orbitals, satisfying the $\Delta j$=$\Delta l$=3 relation, are placed close to the Fermi level. One member of the $\Delta j$=$\Delta l$=3 pair is a high-j intruder orbital. Such orbitals play an important role in generating spin of fast rotating nuclei and give rise to the characteristic backbending phenomenon, observed commonly in nuclei with quadrupole deformation. An octupole interaction provides an extra force binding the intruder orbital to the, so called, normal-parity orbitals, changing thus its response to fast nuclear rotation so, that backbending is smoothed and delayed. Observation of spin alignment as a function of rotational frequency in nuclei with strong octupole interactions provides, therefore, a usefull tool to study these interactions [1].

A standard example of such behaviour is that reported for the $^{222}$Th nucleus [2], where instead of a sharp backbending, expected for the proton $i_{13/2}$ intruder orbital [3], a gradual increase of alignment is observed [4].

The early experimental data [2], which reported the alternating-pariy band in $^{222}$Th up to spin I=16, were extended by another measurement [5] up to spin I=24, still showing no sign of any backbending. Calculations [6] assuming the presence of an octupole deformation in $^{222}$Th, reproduced experimental data very well up to the highest observed spin. The same calculations predict that at still higher rotational frequency, around spin I=26 a pair of neutrons in the $j_{15/2}$ intruder orbital will align, displaying a sharp backbending. It has been predicted that an alignment of two high-$j$ pairs should cause a transition to a reflection symmetry in the $^{222}$Th nucleus.

Since the measurement of Ref.[5] had stopped short of spin needed to observe the predicted backbending in $^{222}$Th, a new exeriment has been performed [7], using more efficient $\gamma$ detector EUROGAM1. Surprisingly, no spins higher than observed in Ref.[5] were seen in $^{222}$Th. Intensities of $\gamma$ transitions in the s=+1 alternating-parity band, observed in this new measurement, drop below observation limit at about spin I=26, i.e. where the expected backbending should occur. The maximum spin generated in the $^{208}$Pb + $^{18}$O compound-nucleus reaction used in this measurement is about I=35 [7]. Therefore, the the observed intensity loss already at spin I=26$\hbar$ needs an explanation. The authors of Ref.[7] claim that this decrease is not due to competition of fission, because additional measurements [7] show that the $^{222}$Th compound nucleus is populated at around maximum spin of I=35$\hbar$. An obvious mechanism, suggested in by calculations [6] is that the intensity is drained by the predicted yrast, four quasiparticle band corresponding to a reflection-symmetric configuration, which should appear at spin I=26, after two alignments. Such band was not observed in Ref.[7], however, and the high-spin behaviour of $^{222}$Th remains unexplained.

Rotational frequencies observed in the $^{222}$Th nucleus are relatively low, due to large moment of inertia of this heavy nucleus. It requires population of high spin levels in such a nucleus in order to generate rotational frequency high enough to align a pair of high-$j$ particles. More favourable conditions are encountered in the lanthanide region of strong octupole correlations [8, 9] where, because of lower moments of inertia, nuclei rotate faster than the actininde nuclei of the same spin. Consequently, one may expect here alignment effects at lower spins, which are easier to observe.

## 2. High-spin studies of $^{150}$Sm

When an alternating-parity band has been identified in the $^{150}$Sm nucleus [9, 10] it has been also noticed that a crossing is observed in this band with the positive-parity band, showing features characteristic of a reflection-symmetric configuration. This was thus a similar scenario to that predicted for $^{222}$Th, but here observed at much lower spins. There were also indications that the alternating-parity band continues past the crossing and that both reflection-asymmetric and -symmetric configurations coexist there. These observations were reproduced by calculations [11, 12], indicating that in $^{150}$Sm an octupole deformation is induced by nuclear rotation at spin I$\sim$10 and that at higher spins a reflection-symmetric shape is restored, coexisting with the reflection-asymmetric one. At very high spins, corresponding to rotational frequency $\hbar\omega$=0.40, only the reflection-symmetric minimum in the potential of $^{150}$Sm remains [11, 12]. This prediction is different from the one for $^{222}$Th [6] where both shapes coexist up to very high spins. It was then of considerable interest to verify experimentally these predictions for $^{150}$Sm.

To study high spins in $^{150}$Sm we used, analogously to the $^{222}$Th studies, the ($^{18}$O,4n), compound-nucleus reaction, which populates levels with spins up to I=35$\hbar$. We used a $\sim$3 mg/cm$^2$ thick target made of $^{136}$Xe gas frozen on a lead backing [13]. The cryogenic $^{136}$Xe target was mounted [14] inside the NORDBALL array of Anti-Compton Spectrometers, which was used to measure $\gamma$-$\gamma$ coincidences from the $^{136}$Xe($^{18}$O,4n$\gamma$)$^{150}$Sm reaction at 76 MeV of $^{18}$O beam, delivered by tandem at NBI Risø. In the experiment about 2x10$^8$ $\gamma\gamma$ coincidence events were collected. Analysis of this data revealed the excitation scheme of $^{150}$Sm as shown in Figure 1.

The ground-state, alternating-parity band (s=+1) has been extended up to spin I=22, the S-band up to spin I=28 and a new, negative-parity band has been found. This new band, which crosses the s=+1 band around spin I=20, has been established up to spin I=31. Its properties suggest that it corresponds to a reflection-symmetric configuration. A number of transitions feed the 6106.3 keV and 6448.9 keV levels at the top of the s=+1 band but their intensities are too low for a definite placement of these transitions in the excitation scheme.

The new data confirms the prediction that nuclear rotation enhances octupole correlations in $^{150}$Sm. As can be seen in Fig.2, above spin I=14 the B(E1)/B(E2) branching ratios in the s=+1 band stabilize at the level of 0.7x10$^{-6}$fm$^{-2}$. It is interesting to note that this happens above the crossing with the reflection-symmetric S-band around spin I=14$\hbar$.

A complex pattern of band crossings observed in $^{150}$Sm at medium spins most likely corresponds to the predicted shape change from reflection-asymmetric to reflection-symmetric one [11, 12]. To identify the nature of the observed crossings, two odd-A nuclei next to $^{150}$Sm were studied, the $^{151}$Eu nucleus [14] having an additional proton and the $^{151}$Sm nucleus [15] having one unpaired neutron, relative to $^{150}$Sm. In Fig.3a alignment plots for the positive-parity, yrast band in $^{150}$Sm is compared to similar plots for yrast bands based on the $\pi h_{11/2}$ proton excitation in $^{151}$Eu and the $\nu i_{13/2}$ neutron excitation in $^{151}$Sm, respectively.

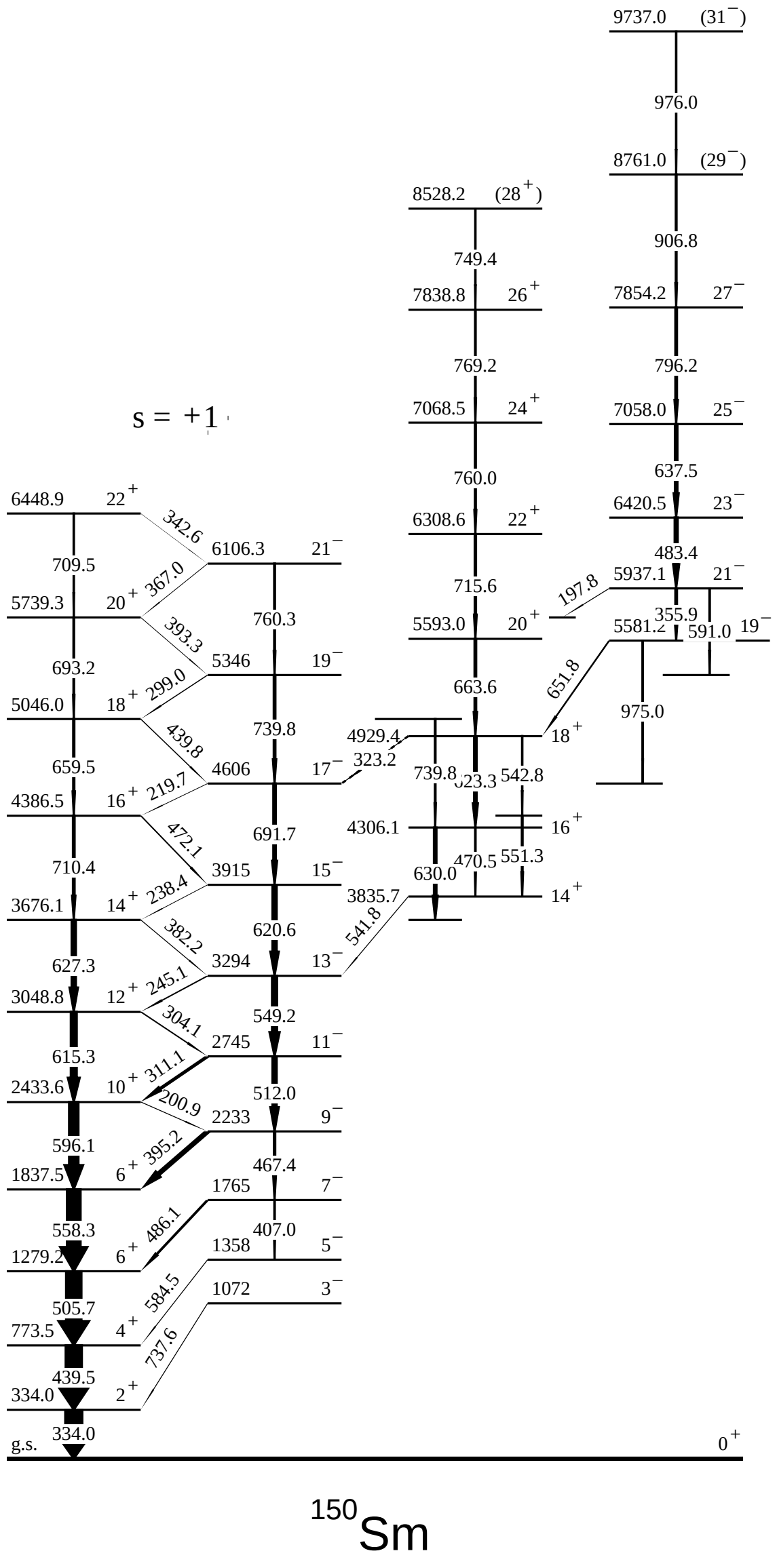

Fig. 1. Partial level scheme of $^{150}$Sm, as obtained in this work.

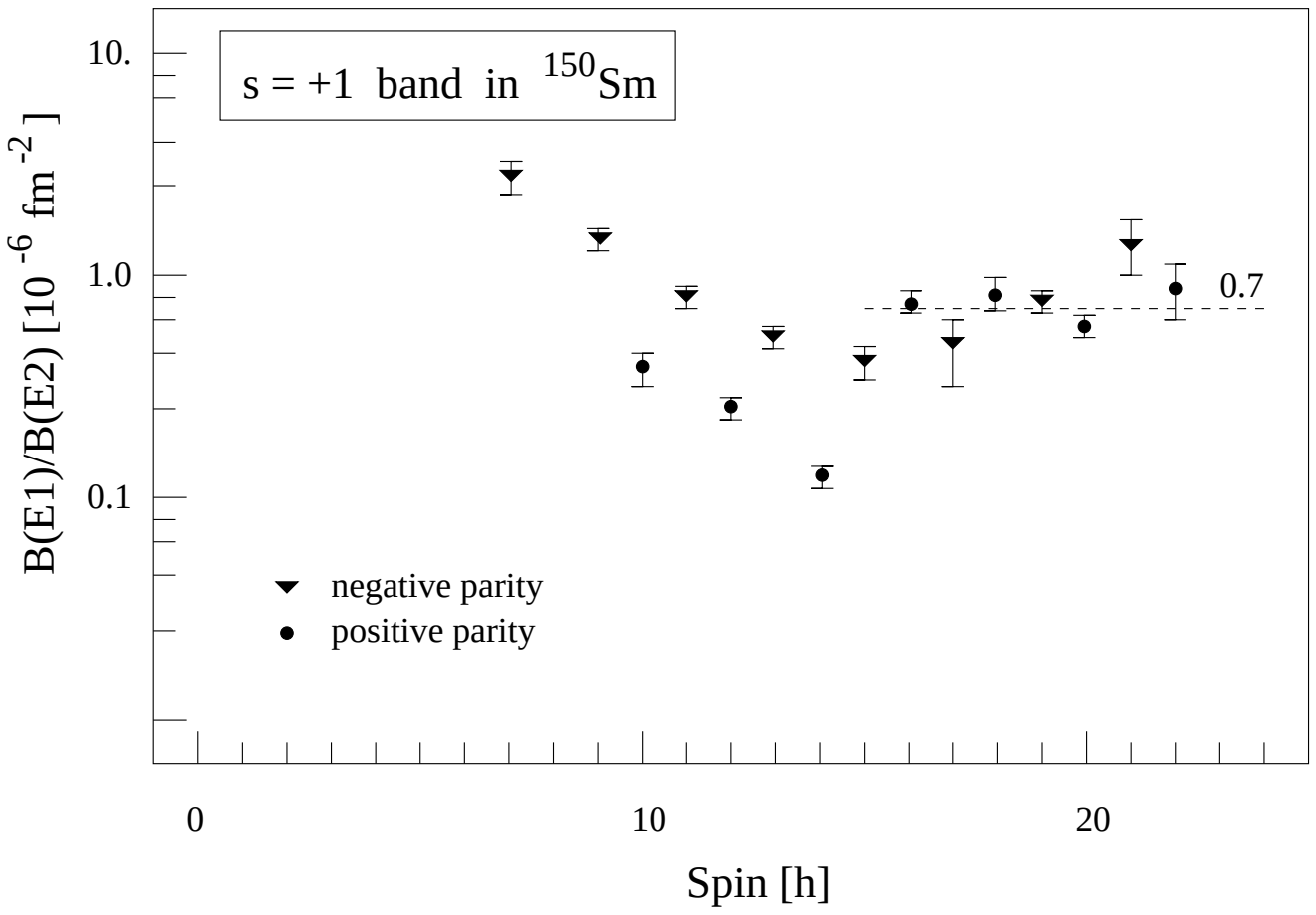

Fig. 2. B(E1)/B(E2) branching ratios in $^{150}$Sm, as observed in the present work.

It is evident from this comparison, that the first crossing observed in the positive-parity band in $^{150}$Sm at around spin I=14, is due to an alignment of a pair of $i_{13/2}$ neutrons since an alignment at similar rotational frequency is observed in the yrast band based on the $h_{11/2}$ proton level in $^{151}$Eu while it is not observed in the band based on the $i_{13/2}$ neutron level in $^{151}$Sm. The frequency of this crossing, $\hbar\omega_c \approx 0.32$ MeV can be compared to the frequency of the AB crossing in $^{150}$Sm predicted at $\hbar\omega_c$=0.23 MeV [16]. The experimental value, which is significantly higher than the predicted one, indicates the presence of strong octupole correlations in $^{150}$Sm, which delay the alignment process in $^{150}$Sm [11, 12].

Further up the positive-parity band in $^{150}$Sm another crossing occurs at a frequency close to $\hbar\omega_c$=0.40 MeV, corresponding to spin I≈24. A crossing at similar frequency starts to show up in the $i_{13/2}$ neutron band in $^{151}$Sm at spin I≈41/2 while no sign of any such crossing is seen in the $h_{11/2}(i^2_{13/2})$ band in $^{151}$Eu up to spin I=43/2. One may therefore conclude that the second crossing in the positive-parity band in $^{150}$Sm corresponds to the alignment of a pair of $h_{11/2}$ protons. Similarly as for the $i_{13/2}$ neutron alignment, the frequency of this $\pi(h^2_{11/2})$ crossing is significantly higher than the predictd value of $\hbar\omega_c$=0.28 MeV [16], indicating again the presence of octupole interactions in the $^{150}$Sm core, which delay this alignment.

Figure 3b. shows alignments in the four bands observed in $^{150}$Sm. In the positive-parity branch of the s=+1 band, the alignment increases gradually. This band is crossed at aroubnd spin I=14 by another positive-parity band, which has alignment $i \approx 12\ \hbar$, a value consistent with the above suggestions

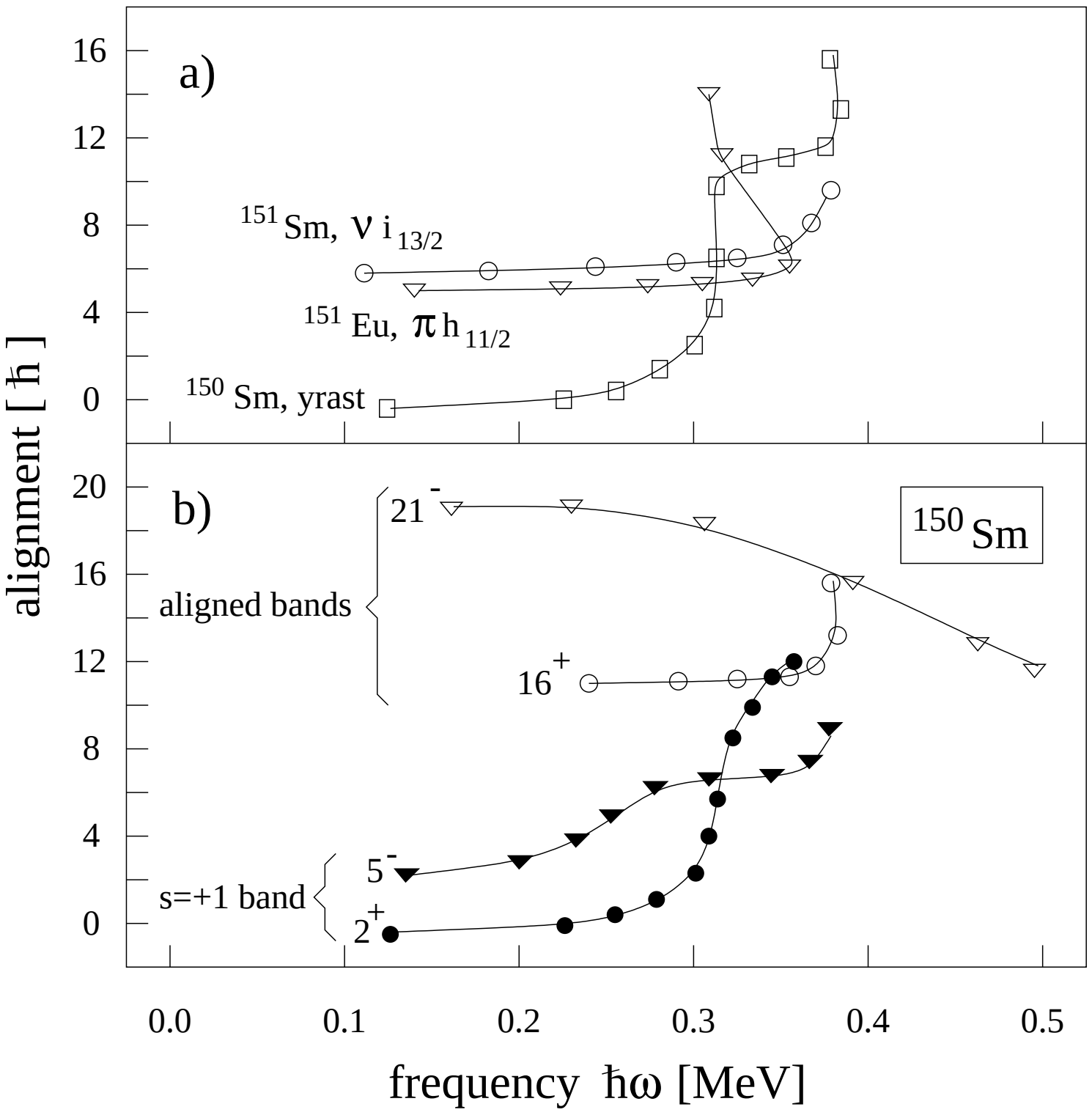

Fig. 3. Alignments in rotational bands in $^{150}$Sm, $^{151}$Sm and$^{151}$Eu. Parameters of Harris plot used were $J_0$=12.0 and $J_1$=130. Lines are drawn to gude the eye.

that this backbending is caused by an alignment of the $\nu(i^2_{13/2})$ pair. In the negative-parity branch of the s=+1 band an uppbend is observed around spin I=11 and the band acquires a moderate alignment of $i{\approx}6$. This phenomenon has been interpreted as due to an alignment in the octupole band [17]. In this case one neutron is promoted to an orbital with higher spin and after that the band corresponds to an aligned, two-quasiparticle configuration. At still higher rotational frequency this band is crossed by the $\pi(h^2_{11/2})$ aligned configuration. The resulting structure has an alignment $i{\approx}20\hbar$, relative to the ground state and probably corresponds to the $pi(h^2_{11/2})nu(h_{9/2}i_{13/2})$, four-quasiparticle configuration. The decrease of the alignment in this band at higher spins, seen in Fig.3b is a result of using the same Harris-plot parameters, as for lower-spin configurations, wherease at higher spins moment of inertia decreases.

## 3. Discussion

The $^{150}$Sm nucleus displays a behaviour similar to that observed for $^{222}$Th, yet here seen to higher rotational frequencies. Studies of the neighbouring odd-A nuclei provide a unique identification of the observed alignment process in $^{150}$Sm. According to calculations [11, 12] tow alignments of the $\nu(i^2_{13/2})$ and $\pi(h^2_{11/2})$ pairs observed in $^{150}$Sm, destroy the octupole minimum in the nuclear potential of this nucleus, where only a reflection-symmetric minimum is left after these two alignments. This is consistent with the sudden drop of intensities of transitions in the s=+1 band in $^{150}$Sm at spin I=22$\hbar$. Figure 4 shows values of these intensities measured in units of their standard deviations, $\sigma$. The observation limit in the present work was adopted at the level of 3 $\sigma$ and is marked in Fig.4 by a solid line. Two aligned bands in $^{150}$Sm, corresponding to a reflection symmetric minimum in the potential, are observed down to the limit of $3\sigma$. On the other hand, the s=+1 band ends at rather high intensities of the last observed transitions. In the negative-parity branch of this band, the highest transition has an intensity of about 40 $\sigma$. Assuming smooth decrease of $\gamma$ intensities in this cascade, one should observe at least two more transitions, which are however not seen. Similar remarks apply to the positive parity branch of the s=+1 band.

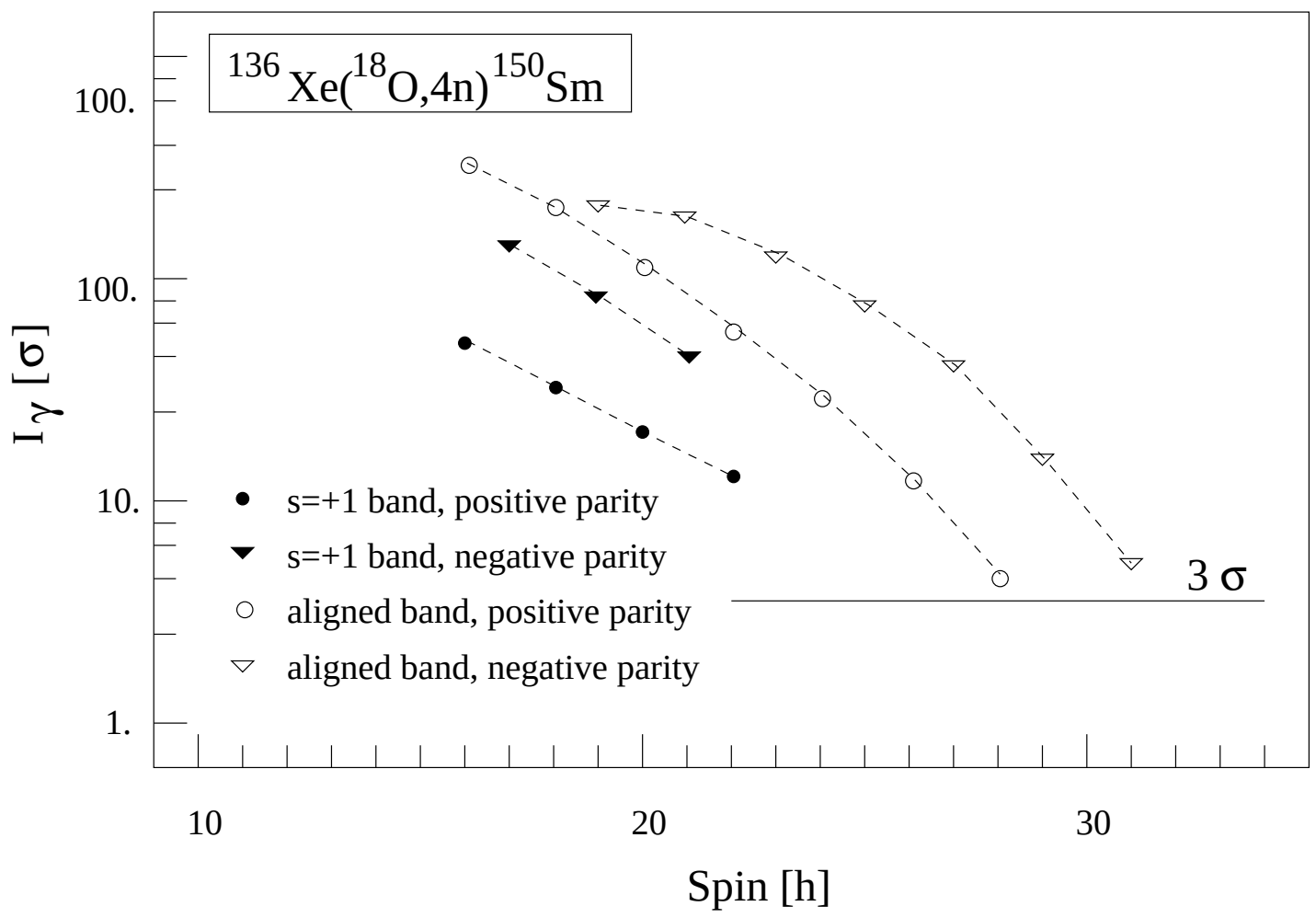

Fig. 4. Intensitis of $\gamma$ transitions in various bands of $^{150}$Sm, measured in units of standard deviation, as observed in the present work. The observation limit of three standard deviations is marked by the solid line. Dashed lines are drawn to guide the eye.

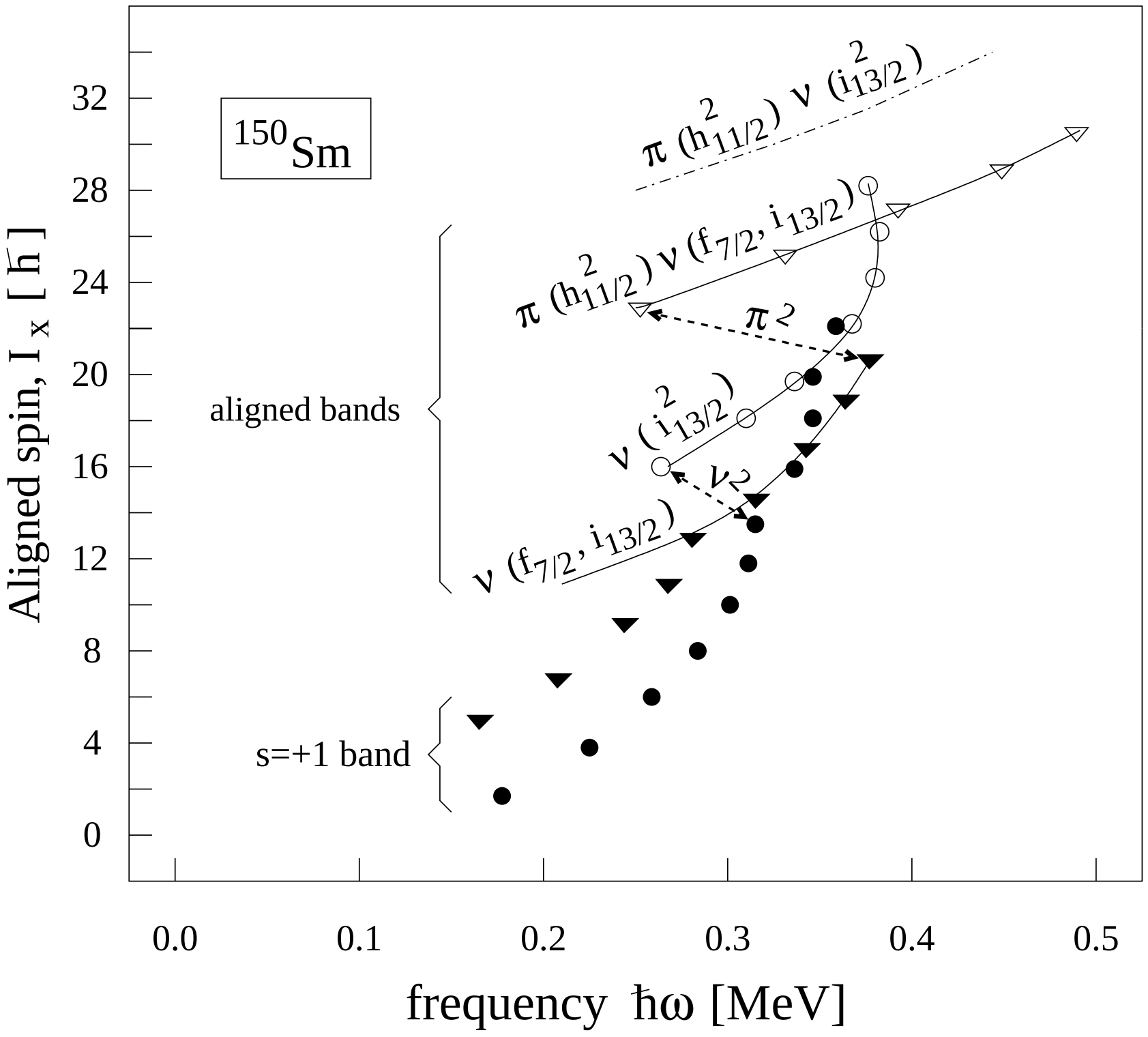

Fig. 5. Aligned angular momentum in $^{150}$Sm as observed in this work

In $^{150}$Sm one has then a more convincing picture of the termination of the s=+1 band than that for $^{222}$Th, available at present. Figure 5, analogous to Fig.12 in Ref.[6], summarizes the experimental evidence of the spin alignment process in $^{150}$Sm. As predicted [11, 12], in $^{150}$Sm both, proton and neutron pairs align in the s=+1 band, causing termination of this band.

The alignment scenario in $^{150}$Sm is somewhat different from the one predicted for $^{222}$Th, where the $\pi(i^2_{13/2})\nu(j^2_{15/2})$, positive-parity configuration should be the lowest one, corresponding to a reflection-symmetric shape, which crosses the s=+1 band. In $^{150}$Sm, where octupole correlation energy is lower than in the actinides, a single $\nu(i^2_{13/2})$ alignment is enough for a reflection-symmetric configuration to emerge, though the s=+1 band still exists past this crossing. It is the next crossing, corresponding to the $\pi(h^2_{11/2})$ alignment, which destroys the octupole minimum in the potential.

The "terminating" 4-q.p. configuration observed in $^{150}$Sm has negative parity, unlike the one one predicted for $^{222}$Th. This is because there are two alignments in the negative-parity branch of the s=+1 band, the mentioned $\pi(h^2_{11/2})$ and the $\nu(f_{7/2}i_{13/2})$ one at around spin I=11. The $\pi(h^2_{11/2})\nu(i^2_{13/2})$ positive-parity configuration, analogous to the $\pi^2\nu^2$ one in $^{222}$Th, is not observed in $^{150}$Sm in this work. It is likely that this configuration is just above the $\pi(h^2_{11/2})\nu(f_{7/2}i_{13/2})$ 4-q.p. one, as marked in Fig.5 by the dot-dashed line. This is suggested by a clear uppbend seen at the top of the $\nu(i^2_{13/2})$ 2-q.p. band in $^{150}$Sm, which most likely corresponds to the $\pi(h^2_{11/2})$ alignment.

As mentioned above, there are $\gamma$ lines feeding the highest levels in the s=+1 band in $^{150}$ but their intensities are too low for their definite placement in the band. It is of a great interest to perform a new measurement for $^{150}$Sm, with statistics much better than the present one. Such experiment could definitely answer the question if the s=+1 band is terminated around spin I=21, as suggested by the present data and the calculations [11, 12]or if there is a continuation of this band, corresponding perhaps to a lower quadrupole deformation. Such a scenario is predicted for $^{222}$Th, where after two alignments, an octupole minimum in the potential appears at nearly zero quadrupole deformation. It is interesting to mention that this kind of band has already been observed [18] in the weakly deformed, $^{218}$Ra nucleus, having 86 protons, two protons less than $^{222}$Th. An interesting feature of this band is that it continues past two subsequent alignments [18]. The reason why we propose to look for such a band in $^{150}$Sm is that similar band has been observed in the N=86, weakly deformed $^{148}$Sm isotope [19]. In this nucleus the s=+1 band is observed up to spin I=27, showing no sign of any abrupt ending, though its structure is very irregular around spin I=20 and it shows a similar pattern of alignments as the s=+1 band in $^{218}$Ra. In this context it is also of considerable interest to perform improved measurements of the Z=86 and Z=88 thorium isotopes, in order to look for the expected, 4-q.p. reflection-symmetric configuration in $^{222}$Th and s=+1 bands of low quadrupole deformation in $^{220}$Th and $^{222}$Th nuclei. We estimate that a repetition of mesurements performed in this work and Ref.[7], using much more efficient array of Anti-Compton Spectrometers, like for instance GAMMASPHERE, should be enough to resolve the probles outlined above.